# Controlling Context: Generative AI at Work in Integrated Circuit Design and Other High-Precision Domains

EMANUEL MOSS, Intel Labs, USA
ELIZABETH WATKINS, Intel Labs, USA
CHRISTOPHER PERSAUD, Intel Labs, USA
PASSANT KARUNARATNE, Intel, USA
DAWN NAFUS, Intel Labs, USA

Generative AI tools have become more prevalent in engineering workflows, particularly through chatbots and code assistants. As the perceived accuracy of these tools improves, questions arise about whether and how those who work in high-precision domains might maintain vigilance for errors, and what other aspects of using such tools might trouble their work. This paper analyzes interviews with hardware and software engineers, and their collaborators, who work in integrated circuit design to identify the role accuracy plays in their use of generative AI tools and what other forms of trouble they face in using such tools. The paper inventories these forms of trouble, which are then mapped to elements of generative AI systems, to conclude that controlling the context of interactions between engineers and the generative AI tools is one of the largest challenges they face. The paper concludes with recommendations for mitigating this form of trouble by increasing the ability to control context interactively.

CCS Concepts: • **Human-centered computing** → **Empirical studies in collaborative and social computing**; *HCI theory, concepts and models*; *Computer supported cooperative work*.

## 1 Introduction

Generative AI tools (GenAI), as exemplified by a range of generative pretrained transformers (GPTs) that began entering the market in November, 2022 [70], have been broadly heralded as transformational tools for productivity and efficiency [121, 75]creativity [108], and the alleviation of tedium [90]. To pursue such promises, individual organizations have begun to implement customized and purpose-specific applications of GenAI as a way to leverage prior computational work (e.g., the production of documents and software code) for the generation of new work products by fine-tuning and augmenting existing so-called "foundation models". Such pursuits expand the scope of how work is accomplished cooperatively through computation by placing workers that are spatially and temporally disconnected from one another into relationships that do not easily fit into existing frameworks for understanding human-computer interactions [3]. As will be explored in greater depth below, the spatial and temporal dislocations represented by foundation models have played a significant role in the host of concerns being raised about the propensity of GenAI to "hallucinate" [82], "bullshit" [112, 94], and "confabulate" [130]. Put less dramatically, GenAI is thought to possibly introduce inaccuracies into existing workflows, and therefore to raise the question of whether or not they can be relied upon for high-stakes cooperative work domains where accuracy and precision are paramount. Additionally, early research indicates that indiscriminate use of GenAI tools in software development teams can increase the number of software flaws as it increase software programmers' coding speed [92] or lead to "deskilling" of programmers and

Authors' Contact Information: Emanuel Moss, emanuel.moss@intel.com, Intel Labs, Hillsboro, Oregon, USA; Elizabeth Watkins, Intel Labs, Hillsboro, Oregon, USA, elizabeth.watkins@intel.com; Christopher Persaud, Intel Labs, Hillsboro, Oregon, USA; Passant Karunaratne, Intel, Santa Clara, California, USA; Dawn Nafus, Intel Labs, Hillsboro, Oregon, USA.

other knowledge-workers through over-reliance on such tools[128]. This tension between the promise of productivity and concerns about inaccuracies is amplified by increasing pressure placed on workers to use such tools, which raises a significant set of questions about whether and to what extent those who work in domains that demand accuracy and precision (e.g., medicine, engineering, scientific research) trust GenAI outputs to be accurate, how they manage inaccuracies, how they might still use GenAI outputs productively even if those outputs are inaccurate or imprecise, and how they might effectively organize efforts to manage the effectiveness of such tools within the context of team- and organization-wide goals [25, 44, 50, 10].

This paper addresses these broad questions by reporting on a series of intensive qualitative interviews (n = 17) with engineers who use GenAI for hardware and software engineering applications related to integrated circuit (IC) design. These interviews were designed to answer our empirical research questions about how work practices were (and were not) changing to ensure desired levels of precision in work products, the extent to which engineers encounter difficulty with the (in)accuracy of GenAI outputs overall, what other difficulties—apart from inaccuracy—engineers encounter when using GenAI tools on the job, and how they recover from troublesome encounters. The paper ultimately concludes that while engineers frequently encounter various forms of "trouble" when using GenAI, their concerns about accuracy were secondary to other, more troublesome aspects of their engagement with this rapidly developing technology. Crucially, all these troublesome aspects are manifestations of the gap [20] between the general-purpose nature of these tools, intended to be used across scales and domains, and the particular context in which they are brought to bear on concrete engineering problems according to the unique conventions of specific organizations, specialties, and teams.

Across every domain of AI development, accuracy metrics are (for better or worse [71, e.g.]) crucial to the task of demonstrating that an AI system has "learned" [89]. Indeed, reducing error for some known function is the key task of machine learning [49], and identifying that function is the key social practice that legitimizes the use of AI across domains [61]. However, the imperatives that drive GenAI development may vary from those that drive GenAI use. Apart from concerns about inaccuracy, engineers reported encountering a wide range of trouble when using GenAI tools. Here, and throughout this paper, "trouble" is taken to mean any difficulty or inconvenience users experience when interacting with a GenAI tool such that they A) must return to the interface and prompt the GenAI to refine its output, B) must edit or alter the output to make use of it, C) can articulate a way in which the output is suboptimal and could potentially be made better-suited to their needs, or D) resort to other, non-GenAI tools to achieve their goals.

In this paper we work to identify forms of "trouble" where the existing literature might otherwise suggest we look for "socio-technical gaps", which Ackerman refers to as a "fundamental mismatch between what is required socially and what we can do technically" [20]. Whereas Ackerman places socio-technical gaps as a "necessary problematic" in cooperative computational work that can be approached from a number of directions (e.g. technical innovation, user education, infrastructural adjustment), we focus here on how these gaps emerge for users and users' strategies for addressing such gaps. This stands in contrast to the type of sustained interrogation that Ackerman calls for in addressing gaps socially and technically. The eruptions of socio-technical gaps into users' awareness are the objects of our analysis, and so we opt towards "staying with the trouble" [46] throughout our analysis, mapping how users recover from and repair trouble, within their interactions with the GenAI tools they use and across the existing workflows the use GenAI within. Here trouble is both a signal produced by human experiences and a property of the uneasy fit between the social and technical marshaled within the systems they use. As a consequence of this focus, we draw developers' attention away from accuracy as the dominant source of trouble, and towards the mixture of pipelines, grounding techniques, and interfaces that places the majority of the burden in overcoming sociotechnical gap on those who use these systems. Below, these troublesome aspects are enumerated in detail, but they circulate around the challenges users face in 'recovering' from and 'repairing' interactions that stem from a fundamental mismatch between the general-purpose scope these tools are developed for and the far more constrained scope users' tasks require.

This paper makes four main contributions to the field(s) of computer-supported cooperative work:

(1) Evidence that concerns engineers might otherwise be expected to have about the accuracy of GenAI is offset by their already existing work practices that prioritize accuracy and precision, which GenAI is gradually being incorporated into.
(2) The introduction of *trouble* as an undesirable feature of GenAI interactions that is orthogonal to concerns about accuracy and inaccuracy, but instead captures the challenges that engineers (and other users of GenAI) encounter when incorporating GenAI into their work practices. This constitutes a novel contribution to the literature focusing on GenAI in workspaces.
(3) A mapping of trouble as it arises through various aspects of GenAI systems, understood as sociotechnical systems with multiple locations for intervention to manage such trouble.
(4) A set of recommendations for specific technical and social interventions that can reduce trouble for engineers employing GenAI on-the-job.

Following a brief review of relevant literature and a explanation of the research methodology employed, this paper will discuss each of the above contributions in turn. The paper will conclude with a discussion of how the findings and recommendations presented here inform future research and provide valuable insights to build on, but extending its findings to other high-precision domains would require further comparative analysis of domain-specific tasks. .

## 2 Literature Review

Because the scope of this paper includes an in-depth analysis of both the technical details of GenAI systems and the implications of those details for a sociotechnical analysis of how GenAI figures in cooperative work, and because the intended audience of this paper includes those who might not be well-versed across technical and nontechnical domains, a review of both the technical and sociotechnical literature is provided here.

### 2.1 Technical Approaches to Generative AI

Generative AI, as the name suggests, is a particular set of artificial intelligence capabilities oriented toward generating seemingly novel outputs across several modalities including text, image, video, and audio. These capabilities differ from other AI capabilities, which are more oriented towards making inferences based on a more constrained set of data used to train models. Such inferences might be used to classify or rank data points that are not part of the original training data by, for example, classifying a image as containing a particular object or ranking the tone or 'sentiment' of a text passage as having some degree of 'toxicity' [80, 64]. For GenAI applications, these capabilities are combined and extended to produce outputs that are not present in the training data, and which satisfy a prompt provided by users. Typically, this is achieved by predicting which words, pixels, or frequencies are most likely to follow (or be situated near) one another based on a mapping of the user's prompt to the vast corpus of training data [48]. This is most clearly illustrated by applications that use generative pretrained transformers (GPTs), which calculate the statistical probability of words and word-sequences that are most likely to follow each other given a prompt [118][1]. Building on such capabilities, GPTs are now being augmented by techniques like retrieval-augmented generation (RAG) [60] to leverage existing documents that organizations produce, to facilitate the generation of organization-specific text, images, computer code, and other outputs. Increasingly, GenAI is becoming a key computational means through which teams working within professional organizations conduct their work [91].

---

[1]This is a simplified description of the precise behavior of GPT-like models, which actually generate likely sequences that are then selected from using other criteria, including instruction tuning [95] and reward models that are trained on the explicit preferences of human crowdworkers [47].

While the capabilities demonstrated by GenAI tools have been impressive, significant concerns accompany their outputs. Text outputs can be correct or appropriate in their form while being incorrect or inaccurate in their meaning [53]. GenAI is trained on which words occur near each other in text scraped from the open web [97] and uses those statistical relationships to produce grammatically (and superficially) valid sentences and paragraphs, but it is not trained on the meaning or correctness of those materials. By way of illustration, the sentences "William I of the Netherlands became king of Holland in 1806", "William I of the Netherlands became king of Holland in 1807", and "King Louis Bonaparte became king of Holland in 1806" might be presented as outputs of a GenAI, are all grammatically valid and plausible to non-experts in 19th century European history. But these sentences are not all factually correct or accurate. This propensity of GenAI to produce plausible but inaccurate outputs has been referred to as "hallucination." Hallucinations can be exacerbated by the quality of data in the training corpora or the instructions (or "metaprompts") provided alongside users' prompts instructing the systems to be helpful (and thereby to return an answer whether or not it can be algorithmically associated with a document in training data) [82].

None of this is to say that GenAI is *incapable* of returning factually correct answers. Numerous tools, tests, and benchmarks have been developed to improve and assess the degree of accuracy of which GenAI systems are capable [85, 127, 120, 84]. However, benchmarking how well a model produces a desired result does adequately capture the many usecases in which the 'correct' answer depends on the context of use. Some of these limitations of benchmarks have been pointed out [116], but in reviewing the literature for this project the preponderance of concerns about model accuracy appeared to overwhelm the much more salient organizational concerns authors had about how outputs could be tailored to specific usecases. These concerns have partially been addressed by RAG-based approaches provide relevant documentation alongside users' prompts from which outputs can be more directly drawn and (importantly) provide links and citations to those documents that may contain accurate facts [83, 68, 60, 63]. These promising advances may greatly *reduce*, but do not *eliminate* inaccuracies in GenAI outputs. Furthermore, significant questions remain about how organizations can best incorporate the beneficial aspects of GenAI while minimizing disruption caused by errors and inaccuracies, while also ensuring that outputs are not just accurate, but also *appropriate* for specific usecases, even when those usecases are not previously anticipated by existing benchmarks.

## 2.2 Sociotechnical Approaches to Generative AI

Questions about how business practices might be inadvertently altered, or might need to be deliberately altered to accommodate GenAI, have recently received attention from researchers using a sociotechnical analytic lens. Attention to the relationship between emerging GenAI technologies and changes to organizations stem from what Wanda Orlikowski et al. refer to as the "metastructuring" effects of technology [14] and raise questions about how GenAI might be accommodated, or stymied, by organizational factors. Such insights from organizational sociology underline the importance of approaching GenAI systems as *sociotechnical* systems, i.e. as systems that have both social components and technical components, neither of which fully determines the functioning and impacts of the whole [16, 55, 62]. Addressing how workers collaborate, with each other and their tools, is crucial to understanding the impact of new computational tools in the workplace. Recent studies of GenAI in the workplace taking a sociotechnical or organizational approach have tended to use a structural lens. For example, these have included studies on how the introduction of GenAI can change roles on teams and the shape of team-based workflows [122, 111, 124], and organizational practices such as project management [99, 76]. Others have built on this approach to explore design implications for GenAI in software development [105].

Addressing how workers collaborate and the impact of new computational tools is particularly important for domains where accuracy and precision have historically been the primary values around which work is organized [7, 39, 51]. Software engineering has long been a focus of scholars examining relationships between individual engineers, wider organizational structures in which they are embedded, and the tools they use [56, 101, 65, 54, 50]. More recently, software engineering has been a site of intense focus for researchers examining how GenAI can be incorporated into work practices, particularly through the development of AI-powered software engineering 'assistants' like Microsoft Co-Pilot (built with OpenAI's GPT technology in the backend), Anthropic's Claude Sonnet, and Mistral's Codestral. Such tools purport to handle a number of complex tasks. These include composing entire applications in any number of programming languages using only plain-language prompts describing a desired functionality, being able to explain what code does within an already-existing program, and aiding in many other software engineering tasks like writing subroutines and functions, debugging, code-commenting, and code optimization [78].

Science and technology studies (STS) discourse on accuracy, maintenance, and repair also has much to offer sociotechnical studies of GenAI. "Accuracy" has been treated variously within STS literatures as socially constructed through negotiated agreements about what constitutes accuracy [37], products of situated practices that put humans and non-human actors into relation in ways that give meaning to concepts like accuracy and precision [4, 19, 31], and as the result of situated decisions that enact epistemological, ontological, and ethical imperatives [34]. "Maintenance and repair" refer in part to behaviors that people engage in to make tools, tasks, and obligation "fit" together in what would be recognized as a workflow, originally referred to by Anselm Strauss as "articulation" labor [5]. The concept of "repair" specifically is attributed to Steven Jackson, in his chapter "Rethinking Repair" [40]. He argues that it is the spontaneous, imaginative labor of human beings "fixing" broken technological systems, to the extent that they can be made "useful", which is truly responsible for any resilience and value which these systems manage to deliver to people. These efforts also echo what Schmidt and Bannon (1992) argue is a central concern to the field of human-computer interaction in understanding the "articulation work" that happens when workers engage in cooperative efforts to re-situate multifaceted complex tasks with different technologies within specific contexts, which they refer to as local "work environments" [11]. The onset of GPT-based tools, combined with managerial encouragement to use these tools, creates an apt environment in which to examine both how locally specific notions of "accuracy" diverge from more global concerns about GenAI tools' performance with respect to specific benchmarks, as well as contextually situated repair practices: how do people "fix" the errors, hallucinations, and confabulations of GenAI output so that such outputs can be truly useful?

Integrated circuit (IC) design and manufacturing, the domain of hardware and software engineers interviewed for this paper, is one such area in which GenAI has been suggested to have tangible impact [86]. In IC design, hardware and software engineers working across multiple teams must precisely describe, analyze, and verify designs using code-like programming languages such as SystemVerilog [33] as well as custom software applications and scripting programs [1]. These tasks have been identified as compelling sites for GenAI application [93, 74, 103, 88], and engineers have been increasingly encouraged to make use of such tools for greater productivity and output quality [87, 79]. This encouragement has been even more forcefully directed at software engineers, and recent work has only begun to assess the impacts of GenAI on hardware and software engineering [107, 119, 93, 117]. In the remainder of the paper, we will discuss our effort to evaluate how early adopters of GenAI for IC design interact with these tools.

## 3 Methodology

We conducted a series of qualitative interviews among employees of a large, multi-national IC design and manufacturing firm. One series of interviews included hardware and software engineers who were documented as intensive users of GenAI tools, defined as using internally-developed GenAI tools to output more than 500,000 tokens over the 30 days prior to selection for inclusion. Other inclusion criteria were that interviewees were based in the Western Hemisphere, and both hardware and software engineers are sampled across a wide range of job roles (e.g., design verification, silicon architect, physical design, electronic design automation (EDA) tool development, enterprise application development, firmware architecture). Interviews explicitly addressed employees' experiences using internally-developed GenAI tools. These tools functioned similarly to publicly available GenAI chatbot tools, using GPT technology on the backend, but run on secure servers ensuring proprietary information can be included in users' prompts without risking data leakage outside the company. These tools also included additional company-specific safety features and useage limitations, and some of the tools incorporated retrieval augmented generation (RAG) [60], giving users access to company-specific documentation as part of their interactions.[2]

A qualitative interview methodology was employed [38] that emphasized a semi-structured interview protocol [36]. Qualitative interviews are ideally suited to gather data about the set of social practices that shape the effects of sociotechnical systems [41, 45, 58, 52], and are therefore about the concerns engineers across job roles might have about accuracy when using GenAI tools. Identifying concerns about accuracy, for example, requires understanding how accuracy and precision are produced by engineers within their social setting, using the tools at their disposal, and subject to "rituals of verification" [17]. Understanding the forms of "trouble" engineers encounter when using GenAI requires understanding what their goals and reward systems are, how those goals are set and met, and who they work with to achieve those goals. And mapping such troubles to concrete locations within sociotechnical systems, and making recommendations to repair such trouble, requires understanding the individual, embodied experience of interacting with such systems and the particular moments in which trouble erupts, and "staying with" it [43]—in this case in the form of open-ended interview protocols that prioritize extemporaneous follow-ups. The interview protocol is included as Appendix A.

A total of 17 interviews were conducted with hardware (n=10) and software (n=7) engineers, all of whom had earned advanced degrees in computer science or electrical engineering and had been in their current job roles for at least one year. Given the relatively recent emergence of generative AI tools, and especially the fact that the internally-developed GenAI tools had only been available for approximately six months prior to the study, all participants had comparable degrees of experience using GenAI on the job. The series of interviews was reviewed by an internal risk and compliance team, which recommended and approved privacy measures to protect interviewees. Interviews were recorded via videoconference software, which also produced automated transcripts of the interview. Recordings and transcripts of each interview were downloaded immediately after each interview and moved to a secure server for storage. Transcripts were post-processed using a custom python script to convert the transcript from a phrase-based format to a turn-based format.[3] Transcripts were first coded openly, using MaxQDA[4] and then grouped thematically [30] according to themes that emerged within and across the categories of analysis. The thematic groups "trouble" and "repair" (Table 2) are the subject of the greater balance of the analysis presented here, but additional thematic groups included as well as common use cases and "judgments" participants passed on the tools and their performance (Table 1), which the interview protocol was intended to illuminate [22].

---

[2]Detailed technical specifications of the tools are withheld for proprietary reasons, but their implementation and functioning is designed to be comparable to off-the-shelf GenAI implementations that were currently on the market during the time of the study.
[3]In the original .vtt file, timecodes and speakers' names were appended every 2-3 seconds. The python code concatenate each speaker's response to produce a turn-based transcript.
[4]https://www.maxqda.com/qualitative-data-analysis-software

## 4 Encountering GenAI On-the-Job

Engineers were interviewed about their use of several GenAI tools developed internally for use by employees. In these interviews, accuracy played a smaller role than expected in matters of concern raised by engineers, however a category of concerns needing repair and covery [11] emerged from these interviews, which we label as *trouble*. Trouble, here is defined as anything needing repair and recovery. Inaccuracies can be seen as a form of trouble, but is treated here as a separate or special category of analysis here because of the role questions about accuracy played in the study design (discussed above). The tools used by engineers took the form of conversational agents that used a chatbot-style interface [69]. Each of these tools were adaptations of commercial-grade foundation models that ensured proprietary information entered into prompts would not leave corporate-leased protected enclaves of vendors' servers or leak to foundation model vendors. Each of these tools were also supplemented with metaprompts [123] to facilitate the use of the tools by various user roles across the company. These tools were also all being used in a testing mode, in which feedback mechanisms (thumbs up/down, star ratings, and free response commenting) within the user interface (UI) enabled constant refinement by the corporate IT team throughout the study period. Some of these tools used RAG to augment data in foundation models with company- or project-specific information. Engineers used these GenAI tools for a number of tasks, not all of which were directly related to hardware and software engineering. Additionally, some tools allowed users to supplement their prompts with files (e.g., specification documents, test cases, product manuals) that could then be interrogated or used as the basis for generated responses.

The most common engineering use cases (see Table 1) served as alternatives to information search (e.g., about a software command or programming language syntax, general information about a specific domain, or to demystify a specific error message) and for the generation of code snippets (e.g., "Write a function that iterates over elements in a tuple"). Other prominent use cases included the generation of scripts for use in a terminal window, summarization or explanation of code, code optimization, and generating documentation to accompany engineering work products. Non-engineering-specific tasks included the generation of meeting summaries from transcripts (.vtt files), drafting emails, changing the tone of emails, assisting in conducting HR tasks like self-evaluations, and translating between languages for communicating with teammates abroad. The trouble that engineers encountered through using GenAI for these purposes is discussed below.

### 4.1 Accuracy and Inaccuracy

This research proceeded from a hypothesis that the accuracy of GenAI tools would be a primary concern for engineers using such tools to produce software and design integrated circuits. In interviews, engineers generally demonstrated unambiguous understandings of what "accuracy" means to them, within their professional practice. They described accuracy as a property of GenAI tools and their outputs, without questioning, critiquing, or otherwise problematizing the concept of accuracy, which simplified discussions about the importance accuracy held in their use of GenAI tools for integrated circuit development. However, in these same interviews, engineers' concerns about other issues eclipsed those about accuracy, without exception. While several (n=7) engineers expressed opinions about the accuracy or inaccuracy of GenAI in the engineering domain, none saw it as a barrier to using the tools or as a reason not to use the tools for engineering tasks. In most instances, concerns about accuracy are orthogonal to the goals they pursue when they use GenAI on the job. One hardware engineer with four years of experience within the company stated, when asked how important the accuracy of GenAI tools is, "I think it's critical. I know AI does hallucinate, but you know that's where I'm having to use my own brain with it, which is, you know, I'm using it as a tool for improvement." This ambivalence toward accuracy stands in stark contrast to the role accuracy plays in *developing* GenAI tools, where it is a central focus of benchmarking.

| Group | Code | Count |
|---|---|---|
| Usecase [124] | meeting summarization | 18 |
| | alternative to search | 17 |
| | coding | 11 |
| | scripting | 8 |
| | debugging | 7 |
| | optimize code | 6 |
| | summarization | 6 |
| | understanding code | 6 |
| | drafting emails | 5 |
| | ideation / extending knowledge | 5 |
| | not useful | 5 |
| | generating examples | 4 |
| | numerical task | 4 |
| | question-answering | 3 |
| | pattern description | 3 |
| | learning | 3 |
| | generating documentation | 2 |
| | proofreading / syntax checking | 2 |
| | tutorial | 1 |
| | generating tutorials | 1 |
| | OKR / HR tasks | 1 |
| | confirmation | 1 |
| | translation | 1 |
| | organizing information | 1 |
| | compare tables | 1 |
| | decoding acronyms | 1 |
| | paper writing | 1 |
| Judgement [59] | inaccuracy | 19 |
| | stakes of trouble/inaccuracy | 13 |
| | context | 6 |
| | trust | 5 |
| | tradeoff | 4 |
| | privacy | 3 |
| | relative performance | 2 |
| | latency | 2 |
| | comparison to human capabilities | 1 |
| | gave correct answer | 1 |
| | non-textual capabilities | 1 |
| | importance of usecase | 1 |
| | epistemology | 1 |

Table 1. Usecases and Judgements

In contrast, hardware and software engineering disciplines have developed a range of tools and practices for ensuring the reliability of software and hardware, which renders concerns about the 'accuracy' of any one piece of code or component of an integrated circuit somewhat moot. For software engineers, reliability engineering is a robust practice with a vast array of techniques to produce dependability and ensure that "latent" and "dormant" errors can be identified throughout the software development lifecycle [15, 32] through robust code review, unit testing, and many other practices. Similarly, in hardware engineering and IC design, system validation and verification [35, 13] represents more than 50% of the overall development effort [27] and these tools, deployed continuously over the development lifecycle, regularly catch errors and inaccuracies in designs. Therefore, it is not GenAI that hardware and software engineers need accuracy from, it is the overall sociotechnical system—the checks and rechecks, the documentation practices, and the testing systems—around the engineers that needs to be oriented toward accuracy, precision, reliability, and dependability. With these systems in place, engineers' use of GenAI succeeds when outputs are constrained or shaped such that they are "good enough" [102] that these systems can be brought to bear, and any 'inaccuracies' are superficial and easily addressed. In turn, interviewees largely considered the generation of large swaths of code or an entire applications to be an inappropriate and largely futile use, as it thwarted or rendered ineffective their existing social practices oriented toward accuracy, precision, and dependability (See Section 4.3).

In this context, the lack of concern engineers evinced in interviews does not read as a null finding. Instead, it indicates that barriers to engineers' GenAI adoption are not contingent on the perfection of GenAI system performance as measured in common accuracy metrics [85]. Crucially, hardware and software engineers experience distinct forms of "trouble" in using GenAI for their on-the-job tasks, apart from inaccuracy. It is to these forms of trouble we now turn; identifying and mitigating trouble emerges as a critical vector in the development of GenAI for engineering tasks.

### 4.2 Trouble

The most common form of trouble reported (by n=11 interviewees) was how a GenAI system made use of documents supplied to it, either directly in a prompt or as part of a RAG implementation. For example, interviewees described how the GenAI would reply with the contents of adjacent cells from the one that 'should' have been the subject of a response, with a response drawn from a footnote rather than the more authoritative text from the main body of a text. The tool could not use the 'context clues' provided by the layout of a table to make valid inferences about its content. The second most common trouble reported (n=9) was from GenAI responses that were "too generic" to be useful: for example responses described general steps one might take to solve a problem rather than provide the solution asked for in context. A third major category of trouble reported by users (n=6) was difficulty with numerical operations, returning arbitrary responses when asked to convert hexadecimal notation to binary, for example, to count the number of bits in a binary string, or to perform seemingly-straightforward arithmetic or algebraic operations.

Other clusters of trouble reported by n > 3 interviewees included explicit references to "hallucination" (n=5) and "tone" (n=4). Hallucination trouble mapped onto the broader literature on AI hallucinations, in which the tool provided an output that seemed appropriate to the context, but lacked some other form of validity [82]. "Tone" trouble actually covered a wide gamut of interviewee complaints, from overall reports of GenAI's tone being cloying, lugubrious, or overly eager-to-please, to being simply inappropriate for the context (e.g. an upbeat tone for a report of bad news, or an overly casual tone for a company-wide email or report to a superior). Additionally, single interviewees reported trouble arising from how the user interface displayed outputs (e.g., line-by-line, word-by word, or in large text blocks that filled the screen), how conversations with a GenAI-powered chatbot became 'canalized' or went "down a rabbit hole" in hot pursuit of an incorrect solution path, how a GenAI refused a reasonable prompt (in the eyes of the interviewee) and instead explained that providing a response would be somehow unsafe, and how often a GenAI tool would provide a response that was bafflingly unaware of the context in which a prompt was given (even when the user was explicit in providing context). Additional instances of trouble are included in Table 2 below.

| Group | Code | Count |
| --- | --- | --- |
| Trouble [85] | parsing tables / document structure | 11 |
| | too generic | 9 |
| | Intel-specific language | 6 |
| | numerical operation | 6 |
| | hallucination | 5 |
| | tone | 4 |
| | meeting attendance error | 3 |
| | inconsistency / variability | 3 |
| | missing depencies in coding | 2 |
| | resists instructions | 2 |
| | trouble | 2 |
| | only provides structure | 2 |
| | text to image | 2 |
| | transparency | 2 |
| | inability to follow instructions | 2 |
| | coding error | 2 |
| | removes functions | 2 |
| | path dependency | 2 |
| | missing or mistaken context | 2 |
| | version error | 1 |
| Repair [60] | atomizing the work | 13 |
| | iteration | 7 |
| | setting context | 5 |
| | visual inspection | 5 |
| | restart | 4 |
| | explicit articulation / request | 4 |
| | paste in errors | 3 |
| | scaffold the outputs | 2 |
| | rephrase the prompt | 2 |
| | generating a script instead of an answer | 2 |
| | independent search | 2 |
| | direct prompt to fix a mistake | 2 |
| | explain code | 1 |
| | dissemination of knowledge | 1 |
| | supply additional documentqation | 1 |
| | transparency of model training | 1 |
| | team checking | 1 |
| | communicate the stakes / shift the persona | 1 |
| | compiling | 1 |
| | insert failure point | 1 |
| | intitialize prompt with detail | 1 |

Table 2. Trouble and Repair

A majority of these types of trouble have to do with a perceived failure of the GenAI tool to provide a contextually appropriate output that is immediately useful to users. Outputs that are "too generic", that misrecognize terms of art or company-specific language, 'hallucinations', mismatches of tone, inconsistent responses, etc. stem from a mismatch between what users expect and what they get from a GenAI tool. Users' expectations are shaped by the context in which they are working, e.g. the version of the programming language they are using, stylistic conventions that characterize their company's codebase, the social norms of how to communicate up or down the managerial chain within their organization, and so on. When a GenAI tool outputs fail to account for—or appear not to account for—that context, it falls to users of these tools to repair that context, either by returning to the tool and adjusting their request or editing the output directly, to recover something useful from the frayed threads of a troubled interaction. Where the trouble consists of straightforward errors or inaccuracies, engineers rely on existing work practices to repair the faulty outputs of GenAI tools. For both error repair and context repair, the effort that comprises repair work represents an externality that needs to be measured and mitigated as part of any claims to GenAI tool effectiveness or productivity improvement.

### 4.3 Repair and Recovery

As has been discussed above, engineers encountered a range of trouble in their use of GenAI, with inaccuracy seldom reported as a dominant form of trouble. Why these things were experienced as trouble had to do with the need to conduct repair work, which, in various ways amounted to attempts at controlling context. Engineers described a set of practices already common in engineering that they use in tandem with (or as an enveloping set of practices around) GenAI. These practices are longstanding techniques used to forestall, repair, and recover from the common (and frequent) errors and misunderstandings that arise in everyday engineering work, developed quite apart from GenAI but inseparable from engineering practice with or without GenAI. While varying in their particulars, engineers broadly described a set of practices that can be described as 1) *Atomizing the Work*, 2) *Iteration*, 3) *Making Explicit*, and 4) the subsequent use of concrete software and hardware *Organizational Workflows*. Notably, engineers describe the utility of many of these practices for repairing and recovering from non-accuracy related concerns as well.

*4.3.1 Organizational Workflows. Organizational workflows* refer to the expected set of pre-existing practices and organizational forms in which engineers' use of GenAI is embedded. These practices have long given order to the social milieux of software and hardware engineering [18], both as professions and as specialized teams within specific organizations. The outputs of GenAI—like the outputs of individual contributors—are subject to practices like code review, visual inspection by engineers who have developed difficult-to-articulate heuristics for 'what looks right or wrong', and unit tests. These tools and practices have been used for decades to manage work, identify and correct errors in code, and navigate the professional demands of engineering. Here, they are the bulwark of confidence engineers relied on when they expressed little concern about the accuracy of GenAI tools; no element of engineering is error-free but, the interlocking practices of engineering are relied upon to sift out, at an organizational level, errors that arise within any single tool, practice, or person. With this bulwark in place, it is not the prospect of putting bugs into production that troubles engineers, as much as the halting, repetitive, off-target, or overly generic interactions with GenAI tools needed to produce outputs that can then be added to their work along the way to code review or unit testing.

*4.3.2 Atomizing the Work. Atomizing the Work* refers to the practice of reducing complex, multi-step, or cumbersome tasks into smaller, more manageable pieces. In engineering more broadly, large tasks are commonly decomposed to allow work to proceed in parallel [23] or to execute complex tasks piece by piece [42, 8]. For GenAI, engineers described how, in the words of one hardware engineer, "rather than ask for complete code" they are more likely to "work on one function at a time or so. Let's work on one function, put that in my IDE [Integrated Development Environment] and then go from there and work on another one." Another, slightly less-experienced hardware engineer stated that "I definitely have been trying to give it less large sections of code... The more information you give it, the more likely it is to do something chaotic or go a little crazy. If you keep it at 10, 15, 20 lines of code that are pretty concise and have a concise function it does a really good job, but if you start giving it 100 lines and saying, hey, can you make this one change? It can start going off the rails a little bit."

For some, this practice may be couched in an implicit understanding of the GenAI tool's overall capabilities, such that "when I upload the code and I asked it to explain things, there are times where it doesn't understand my question, but I'm able to split it up in multiple parts so it's easier to digest that," as one hardware architect put it. But it also may be strategic, to forestall the potential for error that always exists in collaborative endeavors; the same engineer described this as trying "to be very compartmentalized, so I would paste small functions, so I would be able to gauge and test if it's actually doing the correct thing" in the context of the longer piece of code. For this engineer, atomizing the work in this way had become a regular part of their coding practice, to ensure they had oversight for every line of code they used, and to limit the risk of an unnoticed error slipping through their workflow. Another described this as a means of control: "I don't want it to write the project, I just want it to write the different functions and small components of the project. I still maintain control of the overall project, so I think it limits the ... exposure to risk."

*4.3.3 Iteration.* *Iteration* refers to the practice of refining a task or an output over successive versions until it reaches a desired end state. Iteration is common in software engineering, and is an intrinsic component of the popular agile software development methodology [28], but also has long featured in engineering more broadly as a way to elicit stakeholder input and refine engineering specifications [12]. When using GenAI, an engineer might refine a prompt given to a GenAI iteratively, given the output, until reaching a desired goal. "I've learned that you can continue to massage your ask to a point where it gets you better results." Rather than refine specifications, engineers report using far more linguistic cues (as opposed to tweaking actual parameters that produce an output): "There is definitely some art crafting your prompt to make it do what you want. You can even tell it to prioritize certain things, or respond back saying, 'hey, that's great but I didn't like this part. Can you fix that?' and it will go back and rework the code." Following the software engineering principal of modularity [21], iteration is sometimes used in conjunction with *Atomizing the Work*. One software engineer who has been working on data and databases for more than twenty years said, "I basically iterate, put it in, test it out, then go back, and if everything is up to that point, what I'm expecting. Then we go on to the next step." But iteration can also be used by engineers to generate novel approaches they might not have landed on by themselves, describing it as "more of learning that process of, if I do it this way, what does it generate? Does it generate what I expect it to? So OK, let me reword it a little bit and let me specifically tell it."

*4.3.4 Making Explicit.* *Making explicit* refers to the practice of carefully and explicitly manipulating the information provided to a GenAI tool in a prompt to achieve a desired output. This can consist of "prompt engineering" [77], the practice of rewriting a prompt in response to an output that needs repair, or the deployment of a carefully honed and tested prompt that has been demonstrated to produce contextually appropriate results. This involves explicitly including contextual details needed for an appropriate output. For example, an engineer might 'enroll' a GenAI tool as part of a prompt by telling it that it is an integrated circuit design expert, or producing an output for a specific audience. An engineer might also include code examples exemplifying the style conventions of a team or project, to be emulated by the GenAI tool in its output. It can also consist of pointing the GenAI tool to documents that act as a context for the response. In this sense, RAG is a way to make context explicit, as are many of the affordances of AI coding assistants that treat files that are open or flagged within an IDE as relevant to a prompt or AI-generated code completion.

One reason that making context explicit is crucial for engineers is that GenAI tools draw on GPTs that—while supplemented with engineering-specific data—were designed as general-purpose tools that could conceivably answer any question on any topic [70]. From the tool's "point of view," plausible responses to prompts could conceivably draw on any realm of human discourse, unless they are explicitly instructed not to (through development techniques that will be discussed below). Therefore, engineers have found that orienting their prompt within a more constrained domain of knowledge prevents 'misunderstandings' or inapt responses. ‚In the words of one systems validation software engineer, "it needs the context, right? So if you provide the context and then you start asking questions, then it's more specific to what you're asking." They report going so far as to explicitly declare the professional role the GenAI tool ought to 'play', the role they themselves occupied, and what was expected of the tool: "When I first started writing prompts for [their employer's internal GPT chatbot], I was extremely specific. I would say I am a software developer and a hardware validator and I would like you to create a function that would do $x$ constrained by $y$ and I would make my prompt as detailed and informative as I could possibly do it and it was very good," said one front-end software engineer. This particular engineer continued with a more macroscopic "folk" understanding [57] of how making things explicit might work, saying "but what I realized is I don't have to do that on every single prompt and that's why I have so many conversations here with [this GenAI tool]. Once I make that initial prompt, it seems to maintain that initial understanding through all my subsequent prompts." Many publicly available GenAI tools allow conversational threads of conversation to persist, while others provide affordances to 'reset' the conversation. Attempts to control the context of that conversation, whether new or old, are crucial for engineers' management of expectations around GenAI outputs.

## 5 Mapping Trouble

The trouble engineers experience (Table 2) is not a case of user error or evidence of the need for more training in how to use such tools [109]. Rather, it is a way of experiencing Ackerman's socio-technical gap. The gap between design choices and engineers' goals puts the onus on those engineers to set and reset, steer and re-steer, towards the context they want the tools to work within. This onus is perhaps unsurprising given the prioritization of generalizability and scale by developers of GenAI systems [110]. However, locating where those gaps manifest specifically in design choices, and mapping those choices onto trouble as experienced by users, is complicated by the complexity of GenAI tools and the multiplicity of components that are assembled to comprise a user-facing product. This section uses the forms of trouble discussed above as instructive signposts toward which system components can be targeted, to A) reduce the trouble engineers experience and B) support the practices of repair and recovery they employ. Minimizing trouble and supporting repair cannot happen at the overall system level. Instead, reparative practices must target specific components of the GenAI assemblage.

Trouble can be associated with multiple elements because the elements of any sociotechnical system interact and mutually shape each other [16]. It should be noted, however, that this mapping is specific to integrated circuit designers and the GenAI systems they were using at the time of this study. A similarly designed study of other contexts and domains (e.g., medical professionals using a hospital-developed system) might result in a different mapping. It is hoped there would be rough correspondences of trouble and elements across multiple domains, so that some design implications can be generalized.

| Type of Trouble | Pipeline | Features | Grounding |
|---|---|---|---|
| parsing docs | × | | |
| too generic | | × | × |
| misrecognizes infrequent tokens | × | | × |
| numerical operations failed | | | × |
| 'hallucination' | × | × | × |
| tone | × | × | × |
| UI / display | × | × | |
| unresolvable conversation threads | × | × | × |
| inability to follow instructions | | | × |
| unwillingness to follow instructions | | × | × |
| misunderstands or does not grasp context | | | × |
| outdated responses | × | | |
| faulty temporal or spatial reasoning | | | × |
| gives different responses to identical prompts | × | | × |
| 'pads' outputs with unnecessary verbiage | × | × | |
| does not admit ignorance of a topic | × | × | |

Table 3. Mapping trouble onto elements.

## 5.1 Pipeline

GenAI tools are produced through a *pipeline* that connects various computational techniques to produce the tool itself. This pipeline ingests text—for language models this consists of text and tabular data culled from documents, but large multimodal models (LxMs) might also include images, audio, and video—and results in a model that can transform text-based prompts from users into text outputs (or multimedia outputs for LxMs). The pipeline, then, consists of all the steps needed to move from text to model. This includes the datasets and other training corpora that provide the text used to train the model, but also the protocols used to clean and filter those corpora [98]. It also includes the algorithms used to produce statistical models for language in the training data: the zero-shot learning techniques, the transformer architecture, the parameters used to minimize error according to specific objectives, and many other computational techniques [118]. Specifically, the ability to parse documents depends on how inputs are 'tokenized' and whether tabular data is accounted for by a 'tokenizer'. Elsewhere in the pipeline, how polysemous words are embedded within a language model is shaped by their frequency in training datasets, and the frequency with which alternate meanings of such words appear. Cutoff dates for training data also affect whether outputs are current or outdated. The tone of outputs is, in part, affected by designer-supplied metaprompts (such as instructions to 'be helpful') that accompany user inputs and shape outputs, and in part by how user preferences about tone are anticipated by specific design mechanisms [67].

The pipeline also includes techniques for fine-tuning a GenAI tool and tailoring the range of possible outputs to meet the needs of developers. Fine tuning can consist of supplementing the model with additional domain-specific data, for example making a model more suitable for use in solving problems in organic chemistry [125]. But models are also trained to meet social considerations of what a 'good', 'suitable', or 'safe' output might look like. To achieve this, a reinforcement learning model might be trained to match the enacted preferences of developers, online task-workers, or other communities [67], or given explicit rules to follow via a 'constitution' that 'governs' the tool's behavior [66]. Tokenization, fine-tuning, and reinforcement learning represent design choices about how generalizable a model is, collapsing multiple context-specific meanings into single vectors, bounding the context in which particular outputs might be generated in ways that vary from specific contexts of use, or ensuring potential harms are limited across a majority of use cases even if doing so renders a tool useless in specific use cases where such harms are unrealistic.[5]

Increasingly, GenAI tools are being customized within organizations by augmenting the tool with additional data through RAG, which uses such documents to retrieve information that the orginal GPT would not have had access to. Using RAG to augment a GenAI tool promises to make it more responsive for tasks that relate to the context of the organization. However, the fit between A) the documents provided as part of a RAG implementation and B) the tasks the tool is intended to accomplish is not always straightforward, nor is it always readily apparent to end users. Additionally, organizations can contain multitudes, and local contexts within teams or branches may differ significantly from that of the broader organization-wide context. All of these are components of the development pipeline that produces a GenAI tool, and each can present a source of trouble for engineers using such tools.

### 5.2 Features

For the entire class of GenAI tools discussed here, the model produced through the pipeline described above is made accessible to end users through a front-end interface that shapes their experiences. This front-end interface may introduce trouble. For example, some engineers interviewed in this study reported trouble arising from the ways outputs were displayed on-screen, particularly when the GenAI tool emulated a typewriter by displaying results one letter or word at a time. The pace of this front-display frustrated engineers (particularly those who reported an awareness that hardware latency issues were not to blame for this pace, and that it seemed to them like a deliberate choice made by product designers). Conversely, engineers using another tool expressed frustration when massive blocks of text were displayed that exceeded a single screen's capacity. Relatedly, GenAI tools may attach "metaprompts" [123] written by tool developers to users' prompts. These metaprompts are text instructions that users do not see, that are passed to the GPT model alongside the user prompt, and may include instructions to 'be helpful', 'don't swear or display toxic behavior', or even 'explain your reasoning at an $8^{th}$ grade reading level'. However, these instructions may sometimes be at odds with the intentions of users.

An engineer attempting to draft an email to a scientific research team, for example, might have difficulty prompting the tool to produce text at a sufficient level of complexity if an unseen, undisclosed metaprompt instruction to produce text written at a grade-school level is always attached to that engineer's best efforts. Such a feature is an (invisible) aspect of the context of the interaction with the GenAI tool that potentially produces trouble for the user. Other features wrapped around a GPT model can also cause trouble for engineers. These might include restrictions on the size of the text entry box for user prompts,[6] the types of files that can be appended to a user's prompt, functionality related to the threading of input-output chains (i.e., conversations) and how a specific tool allows for the restarting of prompt chains. Additionally, metaprompts to 'be helpful' dovetail with the propensity of GenAI tools to provide an answer, any answer, whether there is relevant data in the training set or not. This has been suggested as a key source of hallucinations in GenAI outputs [82], but also produces trouble for engineers who find value in knowing about data voids so that they can resort to other methods for accomplishing a task.

---

[5]As would be the case where any discussion of self-harm is blocked, including discussions about how to help someone at risk of engaging in self-harm.
[6]This might also be a function of the GPT model's capabilities, as various models only function with limited context windows for prompts.

## 5.3 Grounding

Another element of GenAI's sociotechnical system are the semiotic worlds in which the text (and other media) making up GenAI tools' training data, inputs, and outputs, are grounded. Texts, and the tokens that comprise them, are generally arbitrary symbols that are ultimately meaningless unless grounded in the social worlds in which they are produced, hold meaning, and are interpreted [6, 9]. Through a range of clever and effective NLP innovations [73] transformer models produce plausible and useful outputs, and so have managed to largely skirt the grounding problem. However, there is an asymptotic relationship between the capacity of GPTs to emulate semantically appropriate language and full linguistic competence [100]. From a positivist perspective,[7] this is because, as Banerjee et al. point out, simple prompts can always be ambiguous,[8] and training data are always necessarily incomplete and already out-of-date [100]. From a phenomenological perspective, however, this is because semantic meaning is not objective data that stands apart from human practices of sensemaking, which negotiate ambiguity and *produce* meaning through context-laden interactions [59]. Put simply, and uniting both these perspectives, GenAI tools present trouble to engineers because the world is not and cannot be fully represented linguistically, and large language models can perform no better at this task than language itself.[9] However, design affordances that increase the amount or specificity of context with which GenAI tools operate can dramatically resolve this ambiguity, as contextual cues are precisely what allow humans to resolve ambiguity when they interact with each other [29, 24, 2].

## 6 Remediating Trouble

While associating forms of trouble with elements of a GenAI tool (see Table 3) is insufficient for remediating trouble, or for supporting users in addressing trouble on their own, it does point out where interventions may be useful. Generally speaking, responsibility for trouble associated with pipeline elements falls on those who make choices about that pipeline. This points to people who make choices about the training, deployment, and fine-tuning of foundation models. Responsibility for trouble associated with feature elements is more difficult to assign. Choices about some model features are made by the same parties, e.g. vendors, that develop the overall pipeline. Choices about other model features might be made by downstream customers or consumers of vendors' foundation models. Nevertheless, a range of interventions can be recommended to reduce trouble for engineers employing GenAI on-the-job. To make these interventions successful requires understanding where to apply leverage. Understanding GenAI tools as sociotechnical systems helps identify not only the elements of GenAI systems that produce trouble (see Section 5) but also to identify where interventions can be made on behalf of engineers, within an organization that develops and deploys GenAI tools for engineers to use. For the GenAI systems analyzed as part of this study, which used vendor-supplied foundation models which were built upon by developers within the same organization as their users, the opportunities for remediating trouble are fairly broad. However, only parts of the pipeline described are subject to direct intervention from within the organization. The foundation model itself, its training data, how it was tokenized, and many other aspects of fine tuning are "black boxed" within the model weights taken up by the organization and built upon [126], although workarounds for this limitation have been proposed [129].

---

[7]See [26] for thorough discussion of positivist and phenomenological approaches to understanding the role of context in human-computer interaction.
[8]They give an example: "the question 'What is the meaning of lead?' might be wrongly interpreted as a query about the chemical element instead of leadership, depending on the context" [100].
[9]Similarly, and expanding beyond language to a more universal semiotic, large multimodal models (LxMs) can perform no better at representing and interpreting reality that our representational systems (video, audio, etc.) permit, and our representational systems only function as representations of reality by virtue of surfeit of human, socially-negotiated effort.

### 6.1 Organizational Interventions

Some interventions in the trouble engineers experience have to do with how work is distributed and organized. Specifically, if the ability of a RAG-based tool to parse documents causes trouble or if the tool has difficulty parsing documents provided as prompt inputs, it may behoove an organization to adapt its documentation practices to interface more reliably with the way GenAI tools parse such documents. Within the organization this research addressed, industry- and organization-specific documentation of technical specifications plays a very important role in hardware and software engineering, which presents an opportunity to design GenAI pipeline components that can better parse internal documentation, or to revisit documentation practices to be more easily parsed by existing GenAI pipelines. Depending on the organization, this may require a target study, trial-and-error, and/or a thorough review of how changing documentation may alter other business processes. Another possible organizational intervention is to ensure that engineering practices oriented towards accuracy and precision, which reduce the overall risk of inaccurate outputs from GenAI tools (see Section 4.1), do not erode as GenAI tools gain broader adoption. A significant implication of the research presented above is that code review, pair programming, unit testing, etc. retain their importance and may even be more important as GenAI use grows.

Additionally, organizations need to ensure novice engineers have pathways to gain the professional experience needed effectively supervise GenAI tools. Nearly half of engineers (n = 8) interviewed as part of this research project, all of whom were intensive users of GenAI tools and mid-career engineers or more senior, reported a lack of confidence that a novice engineer would be effective in identifying inaccurate outputs from GenAI tools. Additionally, these engineers frequently spoke of the need for engineering experience to be able to guide a chat-based interaction with GenAI tools toward a productive goal, saying "I'm using my experience to look at the output and [ask myself] 'Hey, does that does that make sense?'" For both documentation and professionalization, organizations must inventory their own processes and needs to identify the best ways to adapt to the use of GenAI tools and to maintain the growth of expertise and professionalization for engineers.

### 6.2 Transparency Interventions

A significant portion of trouble reported by engineers had to do with their difficulty in understanding and anticipating the exhibited behavior of the GenAI tools they were using. Engineers consistently were surprised and disappointed by the tools' responses to what seemed like straightforward prompts. In these instances, they reported a great deal of trouble steering tools toward more useful outputs. As reported above (Section 4.3.4), controlling context is a key practice engineers use to repair and recover from troublesome GenAI tool outputs. Their attempts to control context, however, are often guided by intuition and guesswork more than an explicit awareness of the assumptions made by the GenAI tool about the context within which it is working. To aid users, developers and deployers of GenAI tools (at both the Pipeline and Feature levels), could support engineers by providing transparency at multiple points. Engineers could benefit from seeing information about the metaprompting that accompanies their prompts, the features of a prompt chain that persist from one interaction to the next, and feedback on the amount of uncertainty latent in both users' prompts and tools' responses. Internally, the organization addressed by this study has implemented a "prompt store" to share useful prompts, but could also internally document metaprompts that shape system behavior for all users.

Such transparency would need to go beyond existing proposals for greater transparency into foundation models [104] and include other components and features of GenAI tools as well. A full exposition on uncertainty in generative AI tools is beyond the scope of this paper, but uncertainty and stochasticity is a fundamental feature of GenAI, and transparency about the level of uncertainty that pertains to specific interactions can aid users in understanding how to steer outputs toward desired ends. This is supported by early research on the calculation and communication of uncertainty [114, 106, 81, 72, 115], as well as by data collected for this study that highlights the trouble specifically associated with variability and uncertainty.

### 6.3 Technical Interventions

Technical interventions, too, are important for reducing the trouble experienced by engineers who use GenAI. The troubles outlined above (Section 4.2) point toward specific interventions which go beyond the purely technical advances often associated with the development of GenAI. One intervention is into how documents are parsed for RAG implementations and as inputs to user prompts. This is the opposite side of the coin discussed in Section 6.1. Here, thinking about RAG functionality in a different way, as providing additional context clues rather than just relevant grounded data, can be especially productive. The documents that are included in a RAG implementation provide context for the interaction between engineer and GenAI tool. It's true that most RAG setups produce a tool that has access to an organization's data, resulting in a context of use that is narrower than a general-purpose GenAI tool—but most are still very broad and could benefit from a more deliberate pursuit of contextual clues. Finer-grained context control would enable individual engineers or smaller teams to more narrowly circumscribe the documents they want to be 'in context' for a set of tasks they wish to accomplish with GenAI.

Another intervention is the technical work needed to efficiently calculate and communicate uncertainty as a property of GenAI inputs and outputs, as discussed above (Section 6.2). If a GenAI tool could provide information about which elements of an output came with higher uncertainty, this could potentially show users where more context could help. This could support and hasten the repair of context. Further work is needed to demonstrate this possibility. Additionally, illustrating the uncertainty a GenAI tool has about how to parse a user's input might aid in the repair of context through real-time prompt rewriting. Such calculations are a non-trivial challenge to produce and use, although early research shows some promise that uncertainty estimation is possible and useful [113, 96].

A third, and little discussed intervention, is into how reinforcement models shift outputs toward the exhibited preferences of users through a process called reinforcement learning from human feedback (RLHF). Most commonly applied to foundation models within an overall model development pipeline, reinforcement models learn a reward function from human feedback [67]. This feedback is procured through online taskworker platforms. These tasks are oriented toward collecting generic feedback on whether taskworkers prefer one of a pair of ouputs produced in response to a wide range of prompts. These prompts must be quite general for developers to achieve any significant degree of coverage across the vast range of possible use cases their foundation models are developed to satisfy. This leaves a significant amount of latent space between domains, and it is unlikely that any single topic—particularly integrated circuit design—has been adequately exposed to human feedback from expert engineers to support this stage of the development pipeline. The ability to do RLHF within specific professional contexts is potentially a key development that would help repair work within specific contexts of use more seamless and less troublesome for users, particularly engineers. That said, additional work is needed to assess the extent to which the trouble experienced by engineers can be attributed to RLHF and how RLHF practices can be employed to mitigate trouble for engineers using GenAI tools.

## 7 Conclusion

This paper has provided evidence that, while important, the accuracy of GenAI tools is not a significant source of trouble for integrated circuit designers who use such tools for integrated circuit design. Rather, they experience other

Controlling Context: Generative AI at Work in Integrated Circuit Design and Other High-Precision Domains	19

forms of trouble that arise from how GenAI tools are developed, the types of features they are enabled to provide to users, and the challenge of accomplishing engineering tasks through applications of large language models (in contrast to other forms of machine learning). These forms of trouble offer analytical purchase over the "sociotechnical gaps" that have long been the object of study for sociotechnical systems and related approaches [20], and point to how these gaps can be studied through the impacts they have on computer users' work lives. In this study, trouble has been shown to create challenges for engineers, as they recover from outputs that are not germane to the contexts in which they work. Their methods for repairing the context of their work, when ruptured by troublesome outputs from GenAI tools are described above, and avenues for further work for supporting these efforts have been put forward. As GenAI tools become more prevalent in high-precision domains, the need to measure, minimize, and mitigate the effects of troublesome outputs, and to enable greater control of context becomes ever more important. This paper suggests paths forward for these efforts.

## A   Interview Protocol

(1) Context-setting questions:
   - How would you describe your role at the company?
   - What software applications do you use, generally (e.g., productivity software, integrated development environment (IDE), EDAs, web browser(s), CAD software, scripting terminal)?
   - What GenAI tools do you use on-the-job? Of these, which are the most important for you?
   - How often do you use GenAI tools on-the-job?
   - What GenAI tools, if any, do you use outside of work?

(2) General usage questions:
   - Would you please describe how you use GenAI tools, in your role? Could you give some concrete examples?
   - Are there other tasks you use GenAI for?
   - What makes a 'good' output for these tasks?
   - Can you tell me about the worst–or a particularly bad–response you have gotten from these tools? What was bad about it? Were you able to navigate that response to still accomplish your task? How?
   - What, precisely, do you do with outputs from GenAI tools? Do you copy-paste into another application? Do you record outputs somewhere other than the tool interface?

(3) Accuracy questions:
   - How important is accuracy in GenAI outputs, for you? Do you have any prompting practices you employ to ensure accuracy in outputs? What do you look for in an output to judge its accuracy?
   - Can you show me a prompt that you are likely to use? In the output, which bit is the worst for it to potentially have gotten wrong? What are the stakes if it's wrong?
   - Do you have a structured process for fact-checking outputs? Walk me through that.
   - Can you tell me about any examples where something seemed correct on the surface, but was actually incorrect in some way?

(4) Usage strategy questions:
   - Do you have any prompting 'tricks' that you have developed over time? Can you give me some examples?
   - Has your approach to prompting changed over time, as your familiarity with these tools has increased?

(5) 'Big Picture' questions:
   - In general, have these tools gotten better (or worse) over time? How?
   - What are some roadblocks that make these tools more trouble to use than they might otherwise be?
   - Which parts of your job role have changed the most since these tools became available?
   - Are there any parts of your job role you can never imagine using these tools for?